\documentclass[journal]{IEEEtran}
\usepackage[left=.5in,right=.5in,top=.6in,bottom=.6in]{geometry}
\usepackage{amsmath,graphicx,amssymb,amsfonts}
\usepackage{cite}
\usepackage{algorithm}
\usepackage{algorithmic}
\usepackage{caption}
\usepackage{subcaption}
\usepackage{balance}
\usepackage[usenames, dvipsnames]{color}

\begin{document}
	
	\title{UAV-Assisted 3-D Localization for IoT Networks Using a Simple and Efficient TDOA–AOA Estimator}

	\author{Mojtaba Amiri and Rouhollah Amiri
		\thanks{Mojtaba Amiri is with the School of Electrical and Computer Engineering, University of Tehran, Tehran, Iran (e-mail: mojtaba.amiri@ut.ac.ir). Rouhollah Amiri is with Department of Electrical Engineering, Sharif University of Technology, Tehran, Iran, (e-mail: amiri@sharif.edu.)}}
	
	\maketitle
	
	\begin{abstract}
		This letter proposes an algebraic solution for the problem of 3-D source localization utilizing the minimum number of measurements, i.e., one Time Difference of Arrival (TDOA) and one Angle of Arrival (AOA) pair. The proposed method employs a closed-form weighted least squares estimator and enables the positioning using a single ground station and a cooperative UAV relaying the signal. Analytical derivations and simulation results demonstrate effectiveness of the proposed approach, achieving near-optimal performance aligned with the Cramér-Rao Lower Bound (CRLB) under moderate Gaussian noise conditions.
	\end{abstract}

\section{Introduction} \label{sec:1}
\IEEEPARstart{L}{ocalization} plays a fundamental role in the Internet of Things (IoT) ecosystem, enabling context-aware services and supporting a wide range of applications such as intelligent transportation, asset tracking, environmental monitoring, and surveillance. Accurate positioning is essential for coordinating the vast number of interconnected devices in IoT networks, particularly in scenarios where Global Navigation Satellite System (GNSS) signals are unavailable or unreliable. As a result, precise and efficient localization has attracted significant research attention across wireless communications, underwater networks, radars, sonars, and wireless sensor systems \cite{sallouha2024ground,Zhao2021,Esrafilian2020,Liang2022}.

Various localization methods have been proposed, including Time of Arrival (TOA) \cite{wu2024novel}, Time Difference of Arrival (TDOA) \cite{motie2024self}, Time Sum of Arrival (TSOA) \cite{bayat2023elliptic,amiri2017exact}, Angle of Arrival (AOA) \cite{zheng2018exploiting}, and Received Signal Strength (RSS) \cite{shamsian2024toa}. Among these techniques, TDOA is widely favored for its high accuracy, low complexity, and ease of implementation. It estimates the source position by measuring the relative signal arrival times at multiple sensors without requiring clock synchronization with the transmitter. Likewise, AOA-based methods determine the source direction by triangulating angular measurements from multiple sensors \cite{sallouha2024ground,zekavat2019handbook}.

To enhance localization robustness and accuracy in IoT scenarios with limited infrastructure, hybrid approaches combining TDOA and AOA measurements have been proposed \cite{yin2015simple,jia2018target,kazemi2019efficient,zou2024simple,pang2024hybrid}. This combination leverages geometric constraints to compensate for the individual weaknesses of TDOA and AOA, offering two main advantages: (1) it reduces the number of required sensors while maintaining high localization accuracy, and (2) it mitigates ambiguities such as ghost targets arising from TDOA-only methods.

Previous studies have demonstrated closed-form estimators using multiple sensors equipped with both timing and angular measurement capabilities. For example, \cite{yin2015simple} presented a two-sensor 3-D localization method combining one TDOA and two AOA measurements, while \cite{jia2018target} extended this concept using total least squares estimation. Similarly, \cite{kazemi2019efficient} and \cite{zou2024simple} explored multistatic target localization using joint TDOA-AOA measurements. However, these methods rely on at least two angle-capable sensors, which impose hardware constraints due to the requirement of antenna arrays—particularly impractical for lightweight airborne or spaceborne IoT platforms such as UAVs and CubeSats.

In this letter, we propose a simple and efficient algebraic solution for UAV-assisted IoT localization that requires only one ground-based sensor and a single UAV relay. The UAV acts as a cooperative IoT node, leveraging its altitude and mobility to extend coverage and improve localization accuracy \cite{bayat2023advances}. The proposed scheme requires only one TDOA and one AOA measurement pair, significantly reducing hardware and synchronization complexity compared to existing methods. Analytical derivations and simulation results show that the proposed closed-form estimator achieves near–Cramér–Rao Lower Bound (CRLB) performance under moderate Gaussian noise conditions, making it a promising approach for energy-efficient and infrastructure-limited IoT localization.

The rest of the letter is organized as follows: Section \ref{sec:2} formulates the localization problem based on TDOA/AOA measurements. Section \ref{sec:3} introduces a closed-form solution to address the problem. The performance of the proposed estimator is thoroughly analyzed in Section \ref{sec:4}. Section \ref{sec:5} presents numerical simulations to evaluate the accuracy and effectiveness of the proposed method. Finally, Section \ref{sec:6} concludes the letter.

\textit{Notations:} Vectors and matrices are denoted by boldface lowercase and uppercase letters, respectively. The transpose and inverse of a matrix or vector are represented by $(.)^T$ and $(.)^{-1}$, respectively. The operator \text{diag}$(\mathbf{a})$ constructs a diagonal matrix from the vector $\mathbf{a}$. Statistical expectation is expressed as $\mathbb{E}(.)$. The symbol $\mathbf{a}^o$ represents the true value of the noisy vector $\mathbf{a}$. Additionally, the notation $\nabla_{\mathbf{a,b}}$ denotes the partial derivative of $\mathbf{a}$ with respect to $\mathbf{b}$, defined as ${{\nabla }_{\mathbf{a},\mathbf{b}}} = \partial \mathbf{a}/\partial {{\mathbf{b}}^{T}}$.

\begin{figure}[htb]
	\centering
	\centerline{\includegraphics[width=6cm]{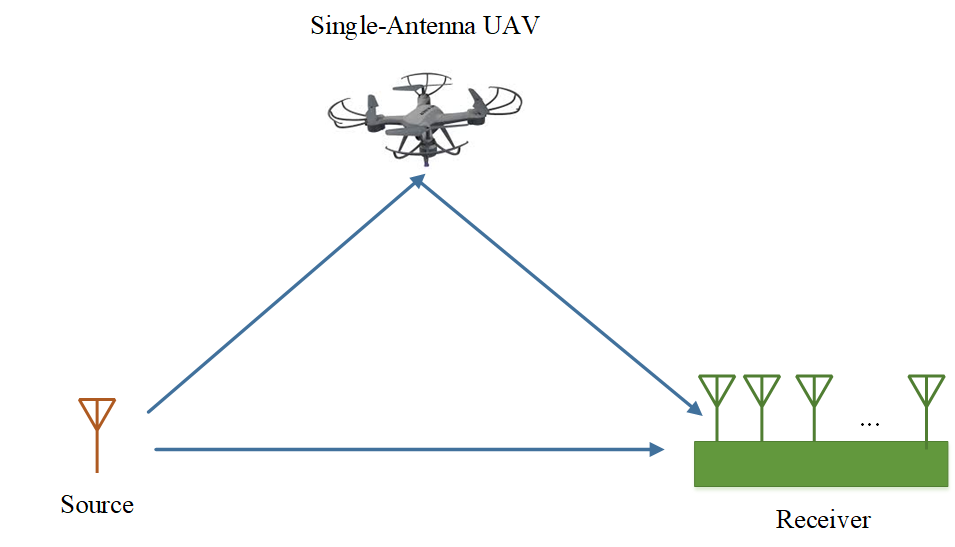}}
	\caption{\small{The localization configuration.}}
	\label{fig:scenario}
\end{figure}

\section{Problem Formulation}\label{sec:2}
This study explores the problem of source localization in a three-dimensional space using a single ground-based sensor, $\mathbf{s}_{1}$ and a UAV stationed at a fixed location, $\mathbf{s}_{2}$. The UAV acts as a relay for the received signals and collaborates with the ground sensor to estimate the position of the source. The primary goal is to determine the unknown source location, denoted by $\mathbf{u}$, by leveraging a combination of TDOA and AOA measurements. The localization framework relies on a set of measurements: the TDOA between the UAV and the ground sensor, and the angular parameters—azimuth and elevation—measured between the source and the ground sensor. These complementary measurements provide the necessary spatial information to accurately estimate the source's position.
The localization scenario is depicted in Fig. \ref{fig:scenario}.

The positions of the ground sensor and UAV are denoted as $\mathbf{s}_{1}=[x_1,y_1,z_1]^{T}$ and $\mathbf{s}_{2}=[x_2,y_2,z_2]^{T}$, respectively.
This system is designed to determine the location of a single target, unknown position of which is represented by ${{\mathbf{u}}^{o}} = {{[ {{x}^{o}},{{y}^{o}},{{z}^{o}} ]}^{T}}$.

The true TDOA at the sensor, when multiplied by the wave propagation speed, is expressed as
\begin{equation} \label{eq: TDOA}
r^{o} = \left\| {\mathbf{u}^{o}}-\mathbf{s}_{2} \right\| - \left\| {\mathbf{u}^{o}}-\mathbf{s}_{1} \right\|
\end{equation}
In this context, the terms range and delay are used interchangeably, as they differ only by a constant factor, namely the wave propagation speed. Due to the presence of measurement noise, the observed value of $r^{o}$ is expressed as $r = r^{o} + \Delta r$, where $\Delta r$ represents the noise component.

The true AOA at the ground sensor is characterized by the pair $(\phi^o, \theta^o)$, where the azimuth angle $\phi^{o} \in \left( -\pi, \pi \right)$ and the elevation angle $\theta^{o} \in \left( 0, \pi/2 \right)$ are defined as follows:
\begin{align}	\label{eq:AOA}
\phi^{o}&\hspace{-.05cm}=\hspace{-.05cm}\text{atan} \hspace{-.1cm}\left( \frac{{{y}^{o}}-{{y}_{1}}}{{{x}^{o}}-{{x}_{1}}} \right), \nonumber\\
\theta^{o}&\hspace{-.05cm}=\hspace{-.05cm}\text{atan}\hspace{-.1cm}\left(\hspace{-.1cm} \frac{{{z}^{o}}-{{z}_{1}}}{\left( {{x}^{o}}-{{x}_{1}} \right)\hspace{-.05cm}\text{cos}\left( \phi^{o} \right)+\left( {{y}^{o}}-{{y}_{1}} \right)\hspace{-.05cm}\text{sin}\left( \phi^{o} \right)} \hspace{-.1cm}\right),
\end{align}
where \(\text{atan}(\cdot)\) denotes the four-quadrant arctangent function. The observed azimuth and elevation angles, \(\phi\) and \(\theta\), are represented as \(\phi = \phi^{o} + \Delta \phi\) and \(\theta = \theta^{o} + \Delta \theta\), where \(\Delta \phi\) and \(\Delta \theta\) are additive noise terms.

By combining the measurements of $r$, $\phi$, and $\theta$, the complete measurement vector is expressed as $\mathbf{m}={{[ {r},{\phi },{\theta}]}^{T}}={{\mathbf{m}}^{o}}+\text{ }\!\!\Delta\!\!\text{ }\mathbf{m}$, where \(\mathbf{m}^{o}\) represents the true values, and \(\Delta \mathbf{m}\) denotes the measurement noise. We assume that the noise vector $\text{ }\!\!\Delta\!\!\text{ }\mathbf{m}={{[ \text{ }\!\!\Delta\!\!\text{ }{r},\text{ }\!\!\Delta\!\!\text{ }{\phi},\text{ }\!\!\Delta\!\!\text{ }{\theta} ]}^{T}}$ follows a zero-mean Gaussian distribution, with the covariance matrix defining the statistical properties of the noise.
\begin{equation}
{{\mathbf{Q}}_{\mathbf{m}}}=\text{blkdiag}\left( {\sigma^2_{r}},{\sigma^2_{\phi}},{\sigma^2_{\theta}} \right),
\end{equation}
where $\sigma^2_r$, $\sigma^2_{\phi}$, and $\sigma^2_{\theta}$ are the noise variances associated with range, azimuth, and elevation measurements, respectively.

The objective of this study is to estimate the unknown parameter $\mathbf{u}^o$ from the measurement vector 
$\mathbf{m}$ with the highest possible accuracy. This task is complicated by the fact that 
$r^o$, $\phi^o$, and $\theta^o$ 
are nonlinear and non-convex functions of $\mathbf{u}^o$ \cite{kay}. To overcome this challenge, we introduce a computationally efficient closed-form solution that approximates the maximum likelihood (ML) estimator under conditions of low noise.

\section{Closed-Form Localization Method} \label{sec:3}
In this section, we develop a closed-form method to estimate the source position, with the objective of achieving performance that approaches the CRLB. We begin by deriving a pseudo-linear range difference equation through appropriate nonlinear transformations together with the AOA measurements. Subsequently, we formulate the AOA equation and combine it with the TDOA equation, thereby improving the precision of the source position estimate.

The differential range equation (\ref{eq: TDOA}) is reformulated as:
\begin{equation} 
 r^{o} + \| {\mathbf{u}^{o}}-\mathbf{s}_{1} \|= \| {\mathbf{u}^{o}}-\mathbf{s}_{2}\| 
\end{equation}
and squaring both sides yields
\begin{equation} \label{eq:BR}
{r^{o}}^{2} + {{\left\| {{\mathbf{s}}_{1}} \right\|}^{2}} - {{\left\| {{\mathbf{s}}_{2}} \right\|}^{2}} + 2r^{o}\left\| {\mathbf{u}^{o}}-\mathbf{s}_{1} \right\| + 2{\mathbf{u}^{o}}^{T}({{\mathbf{s}}_{2}}-{{\mathbf{s}}_{1}})=0.
\end{equation}

Equation (\ref{eq:BR}) is pseudo-linear with respect to $\mathbf{u}^o$.

A commonly employed method for localization involves the use of multi-level Weighted Least Squares (WLS) estimators to estimate  $\mathbf{u}^o$ \cite{chan1994simple}. Traditional approaches typically estimate both the nuisance parameters and the target position simultaneously. In contrast, the present study introduces an alternative approach that utilizes AOA data to eliminate the nuisance parameters from (\ref{eq:BR}), allowing for the estimation of $\mathbf{u}^o$ in a single stage.
According to the localization geometry, we have
\begin{equation} \label{eq:geometry}
{{\mathbf{u}}^{o}}-{{\mathbf{s}}_{1}}=\| {\mathbf{u}^{o}}-\mathbf{s}_{1} \| \mathbf{d}^{o},
\end{equation}
where $\mathbf{d}^o$ is a unit-norm vector denoted by $\mathbf{d}^{o}= [\text{cos}\left( \theta^{o} \right)\text{cos}\left( \phi^{o} \right),\text{cos}\left( \theta^{o} \right)\text{sin}\left( \phi^{o} \right),\text{sin}\left( \theta^{o} \right) ]^{T}$. Multiplying both sides of (\ref{eq:BR}) by $\mathbf{d}^{oT} \mathbf{d}^o=1 $ gives
\begin{equation}
{r^{o}}^{2} + {{\left\| {{\mathbf{s}}_{1}} \right\|}^{2}} - {{\left\| {{\mathbf{s}}_{2}} \right\|}^{2}} + 2r^{o}( {\mathbf{u}^{o}}-\mathbf{s}_{1})^{T}\mathbf{d}^o + 2{\mathbf{u}^{o}}^{T}({{\mathbf{s}}_{2}}-{{\mathbf{s}}_{1}})=0.
\end{equation}
Therefore, it can be written as follows:
\begin{align}
h_{r}&={{\mathbf{g}}_r^T}{{\mathbf{u}}^{o}}, \nonumber\\
{{{h}_r}}&=r{{^{o}}^{2}}+{{\left\| {{\mathbf{s}}_{1}} \right\|}^{2}}-{{\left\| {{\mathbf{s}}_{2}} \right\|}^{2}}-2r^{o}\mathbf{d}{{^{o}}^{T}}{{\mathbf{s}}_{1}}, \nonumber\\
{{ {{\mathbf{g}}_r}}}&=2[ \mathbf{s}_{1}-\mathbf{s}_{2}-r^{o}\mathbf{d}{{^{o}}} ].
\end{align}
Next, we focus on deriving two AOA equations. By taking the tangent of both sides of the azimuth and elevation equations in (\ref{eq:AOA}), expressing tan$(.)$, and performing cross-multiplication, we obtain the following expressions \cite{wang2015asymptotically}:
\begin{subequations}
\begin{align}
& \mathbf{\boldsymbol{\alpha} }{^{T}}{{\mathbf{s}}_{1}}-\boldsymbol{\alpha}{{}^{T}}{{\mathbf{u}}^{o}}=0, \label{eq: AOA_azimuth_eq}\\
& \boldsymbol{{\beta }}{{}^{T}}{{\mathbf{s}}_{1}}-\boldsymbol{{\beta }}{{}^{T}}{{\mathbf{u}}^{o}}=0, \label{eq: AOA_elevation_eq}
\end{align}
\end{subequations}
where $\!\boldsymbol{\alpha}\!\!=\!\!{{\left[\text{sin}\left( \phi^{o} \right),-\text{cos}\left( \phi^{o} \right),0 \right]}^{T}}$ and $\boldsymbol{\beta}=[\text{sin}\left( \theta^{o} \right)$
$\text{cos}\left( \phi^{o} \right),\text{sin}\left( \theta^{o} \right)\text{sin}\left( \phi^{o} \right),-\text{cos}\left( \theta ^{o} \right)] ^{T}$.
Therefore, we have
\begin{align*}
h_{\phi} = \boldsymbol{\alpha}{{}^{T}}\mathbf{ u}^o,
\end{align*}
\begin{equation}
h_{\theta} = \boldsymbol{{\beta }}{{}^{T}}\mathbf{ u}^o,
\end{equation}
where the $h_{\phi}$ and $h_{\theta}$ are $\mathbf{\boldsymbol{\alpha} }{{}^{T}}{{\mathbf{s}}_{1}}$ and $\boldsymbol{{\beta }}{{}^{T}}{{\mathbf{s}}_{1}}$, respectively.
Stacking the range and AOA equations results in the following formulation:
\begin{equation} \label{eq:12}
\mathbf{h}= \mathbf{G}{{\mathbf{u}}^{o}},
\end{equation}
where $\mathbf{h}={{[h_{r}, h_{\phi}, h_{\theta}]}^{T}}$ and $\mathbf{G}={{[ \mathbf{g}_{{r}},\mathbf{\boldsymbol{\alpha} }{{}},\boldsymbol{{\beta }}{{}} ]}^{T}}$.

Since the noise-free parameters in $\mathbf{h}$ and  $\mathbf{G}$  are not directly available, we replace them with their noisy counterparts, i.e., $\tilde{\mathbf{h}}$ and $\tilde{\mathbf{G}}$. The discrepancy between the two sides is then captured by $\boldsymbol{\varepsilon}$, leading to the following equation:
\begin{equation} \label{eq:13}
\boldsymbol{{\varepsilon }}=\mathbf{\tilde{h}}-\mathbf{\tilde{G}}{{\mathbf{u}}^{o}}.
\end{equation}
The WLS solution of (\ref{eq:13}) can be obtained as \cite{kay}
\begin{equation} \label{eq:14}
\mathbf{\hat{u}}={{\left( {{{\mathbf{\tilde{G}}}}^{T}}\mathbf{W\tilde{G}} \right)}^{-1}}{{\mathbf{\tilde{G}}}^{T}}\mathbf{W\tilde{h}},
\end{equation}
where $\mathbf{W}={\left( \mathbb{E}\left\{ \boldsymbol{\varepsilon}\boldsymbol{\varepsilon}^{T} \right\} \right)}^{-1}$. Due to nonlinearity of $\boldsymbol{\varepsilon}$, it is difficult to form the weighting matrix $\mathbf{W}$ in general. Approximating $\boldsymbol{\varepsilon}$ up to the linear noise terms gives
\begin{equation} \label{eq:15}
\boldsymbol{\varepsilon} \approx \mathbf{B}\Delta \mathbf{m},
\end{equation}
where
\begin{align*}
& \mathbf{B}=\text{diag}(B_r,B_\phi,B_\theta), \\
& {{B}_{r}}=2(r-\mathbf{s}_{1}^T\mathbf{d}{{^{o}}} -\mathbf{d}{{^{o}}^{T}}{\mathbf{\hat{u}}}), \\
& {{B}_{\phi}}=\left\| {\mathbf{\hat{u}}}-\mathbf{s}_{1} \right\|\text{cos}\left( \theta \right),\\
& {{B}_{\theta}}=\left\| {\mathbf{\hat{u}}}-\mathbf{s}_{1} \right\|.
\end{align*}

Based on the distribution of $\Delta \mathbf{m}$, it can be deduced that $\boldsymbol{\varepsilon}$ in equation (\ref{eq:15}) follows a zero-mean Gaussian distribution, with a covariance matrix given by $\mathbf{BQ_m} \mathbf{B}^T$. As a result, the weighting matrix is derived as:
\begin{equation}
 \mathbf{W}=(\mathbf{B Q_m} \mathbf{B}^T )^{-1}
\end{equation}
It is noteworthy that the weighting matrix depends on the unknown target location through $\mathbf{B}$. Therefore, an initial estimate of the target location can be obtained by employing $\mathbf{W}=\mathbf{Q}_{\mathbf{m}}^{-1}$ in the first iteration. This initial solution is then used to update $\mathbf{B}$, allowing for the construction of a more accurate weighting matrix to derive the final solution.

By expressing $\mathbf{u}^o$ as ${{( {{{\mathbf{\tilde{G}}}}^{T}}\mathbf{W\tilde{G}} )}^{-1}}{{\mathbf{\tilde{G}}}^{T}}\mathbf{W\tilde{G}}{{\mathbf{u}}^{o}}$ , we can define $\Delta\mathbf{u}$ as follows:
\begin{equation}
	\mathbf{\Delta u}=\mathbf{u}-{{\mathbf{u}}^{o}}={{\left( {{{\mathbf{\tilde{G}}}}^{T}}\mathbf{W\tilde{G}} \right)}^{-1}}{{\mathbf{\tilde{G}}}^{T}}\mathbf{W}\boldsymbol{\varepsilon} .
\end{equation}

By omitting the higher-order noise terms in $\Delta \mathbf{u}$ and computing the expectation, we approximate $\mathbb{E}({\mathbf{u}}) \approx \mathbf{u}^o$. Thus, the proposed estimator is unbiased, and its covariance matrix is \cite{kay}:
\begin{equation}
	\text{cov} (\mathbf{u})=\mathbb{E}\left\{ \Delta\mathbf{u}{{\Delta\mathbf{ u}}^{T}} \right\}\approx {{\left( {{\mathbf{G}}^{T}}\mathbf{WG} \right)}^{-1}},
\end{equation}
where $\mathbf{G}$ is the noise-free version of $\tilde{\mathbf{G}}$. Substituting $\mathbf{W}$ from its definition, we have
\begin{equation} \label{eq: 18}
	\text{cov}\left( \mathbf{u} \right)\approx {{\left( {{\mathbf{L}}^{T}}\mathbf{Q}_{\mathbf{m}}^{-1}\mathbf{L} \right)}^{-1}},
\end{equation}
where $\mathbf{L}=\mathbf{B}^{-1} \mathbf{G}$.

\begin{figure}[h]
	\centering
	\centerline{\includegraphics[width=7cm]{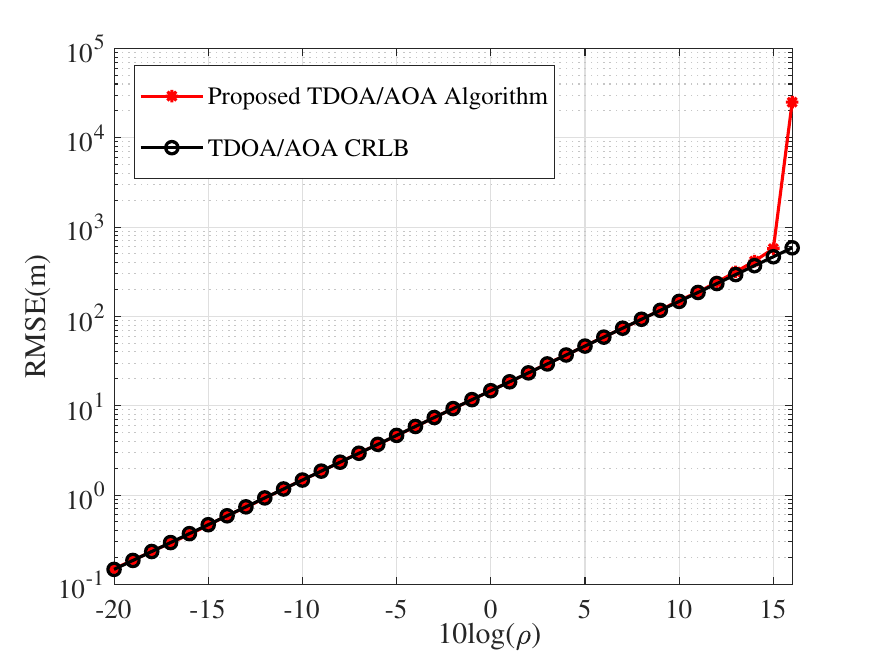}}
	\caption{\small{The performance evaluation of the presented method by comparison with the TDOA/AOA CRLB versus a function of noise scaling factor.}}
	\label{fig:RMSE_SNR}
\end{figure}

\section{Performance Analysis} \label{sec:4}
The CRLB defines theoretical limit on the performance of unbiased estimators \cite{kay} and is used here as a benchmark for evaluation. In the context of the localization problem, the target of estimation is the unknown parameter $\mathbf{u}^o$. Assuming a Gaussian noise model, the CRLB for the estimation of  $\mathbf{u}^o$ is given by \cite{kay}:
\begin{equation} \label{eq:19}
	\text{CRLB}\left( {{\mathbf{u}}^{o}} \right)={{\left( \nabla _{{{\mathbf{m}}^{o}},{{\mathbf{u}}^{o}}}^{T}\mathbf{Q}_{\mathbf{m}}^{-1}{{\nabla }_{{{\mathbf{m}}^{o}},{{\mathbf{u}}^{o}}}} \right)}^{-1}},
\end{equation}
where ${{\nabla }_{{{\mathbf{m}}^{\text{o}}},{{\mathbf{u}}^{\text{o}}}}}={{[ \nabla _{{{r}^{\text{o}}},{{\mathbf{u}}^{\text{o}}}}^{T},\nabla _{{{\phi }^{\text{o}}},{{\mathbf{u}}^{\text{o}}}}^{T},\nabla _{{{\theta}^{\text{o}}},{{\mathbf{u}}^{\text{o}}}}^{T} ]}^{T}}$. The ${{\nabla }_{{{r}^{\text{o}}},{{\mathbf{u}}^{\text{o}}}}}$, ${{\nabla }_{{{\mathbf{\phi}}^{\text{o}}},{{\mathbf{u}}^{\text{o}}}}}$, and ${{\nabla }_{{{\mathbf{\theta}}^{\text{o}}},{{\mathbf{u}}^{\text{o}}}}}$ are given by
\begin{align}
& {{{{\nabla }_{{{r}^{\text{o}}},{{\mathbf{u}}^{\text{o}}}}}}}= \frac{{{({{\mathbf{u}}^{\text{o}}}-\mathbf{s}_2)}}}{\|{{\mathbf{u}}^{\text{o}}}-\mathbf{s}_{2}\|}
-\frac{{{({{\mathbf{u}}^{\text{o}}}-\mathbf{s}_{1})}}}{\|{{\mathbf{u}}^{\text{o}}}-\mathbf{s}_{1}\|}, \nonumber\\
& {{{{\nabla }_{{{\phi}^{\text{o}}},{{\mathbf{u}}^{\text{o}}}}}}}=\frac{1}{{\|{{\mathbf{u}}^{\text{o}}}-\mathbf{s}_{1}\|}\text{cos}\left( \theta^{o} \right)}\mathbf{\boldsymbol{\alpha}} {{}^{T}}, \nonumber\\
& {{{{\nabla }_{{{\theta}^{\text{o}}},{{\mathbf{u}}^{\text{o}}}}} }}=\frac{-1}{{\|{{\mathbf{u}}^{\text{o}}}-\mathbf{s}_{1}\|}}\boldsymbol{\beta} {{}^{T}},
\end{align}
Through direct matrix operations, $\mathbf{L}$ in (\ref{eq: 18}) is simplified, and it can be shown that, under the low-noise conditions, the following relationship holds:
\begin{equation}
\mathbf{L}\hspace*{-.05cm}\approx\hspace*{-.05cm} {{\nabla }_{{{\mathbf{m}}^{\text{o}}}\hspace*{-.03cm},\hspace*{-.03cm}{{\mathbf{u}}^{\text{o}}}}}.
\end{equation}
Consequently, we conclude that:
\begin{equation}
	\text{cov(}\mathbf{u}\text{)}\approx \text{CRLB}\left( {{\mathbf{u}}^{o}} \right),
\end{equation}
when the noise level is sufficiently low.

\section{Simulation Results} \label{sec:5}
In this section, the performance of the proposed closed-form method is assessed through numerical simulations, validating the theoretical findings.

In simulations, the measurement covariance matrix is assumed to take the form ${{\mathbf{Q}}_{\mathbf{m}}} = \text{blkdiag}(\sigma_r^2, \sigma_\phi^2, \sigma_\theta^2)$. The localization accuracy is quantified using the root mean square error (RMSE), defined as $\text{RMSE}(\mathbf{u}) = \sqrt{\sum_{l=1}^{L} | \mathbf{u}^{(l)} - \mathbf{u}^o |^2 / L}$, where $\mathbf{u}^{(l)}$ denotes the estimate of $\mathbf{u}^o$ in the $l$-th ensemble. Each simulation is conducted over 10,000 ensemble runs ($L=10,000$).
The positions of the ground-based sensor and the UAV are set to $\mathbf{s}_{1} = [0, 0, 0]^T$ m and $\mathbf{s}_{2} = [500, 100, 2000]^T$ m, respectively. Three simulation scenarios are considered.

\begin{figure}[h]
	\centering
	\centerline{\includegraphics[width=7cm]{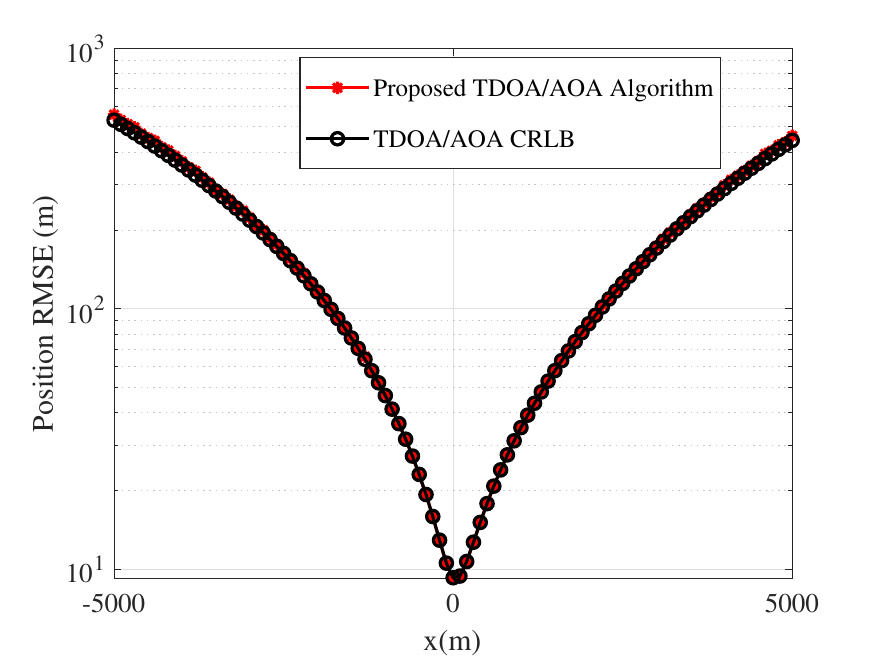}}
	\caption{\small{The performance evaluation of the presented method by comparison with the CRLB for different target positions.}}
	\label{fig:Target_sweep}
\end{figure}

\begin{figure}[h]
	\centering
	\centerline{\includegraphics[width=7cm]{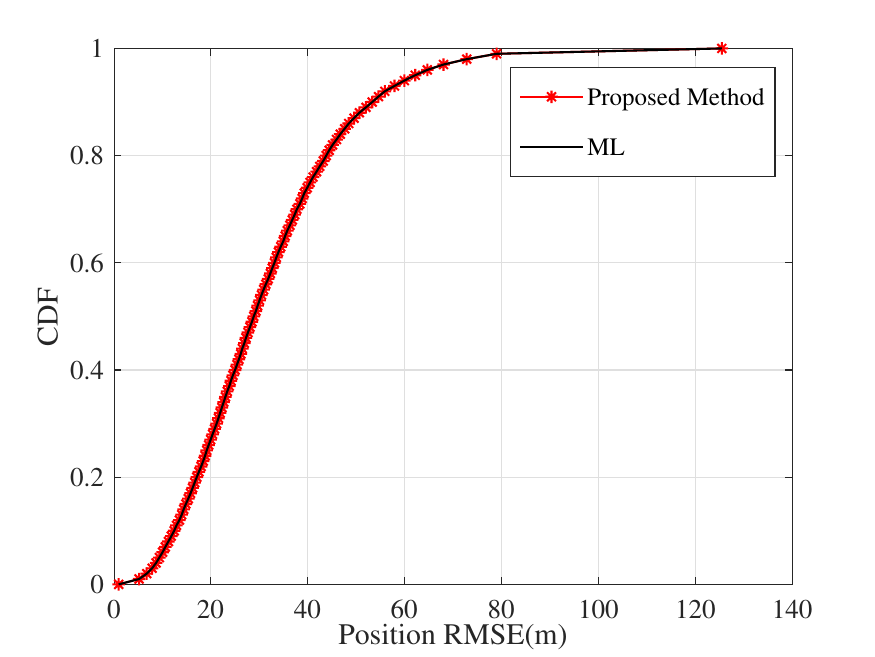}}
	\caption{\small{The CDF of the localization error comparing the presented method with the ML estimator.}}
	\label{fig:CDF}
\end{figure}

In the first  scenario, the target is positioned at $\mathbf{u}^o = [1000, 200, 100]^T$ m. To assess the performance of the proposed method under varying TDOA and AOA measurement noise levels, the noise parameters are defined as $\sigma_r = 40\rho$ and $\sigma_\phi = \sigma_\theta = 0.1\rho$, where $\rho$ serves as a scaling factor.
Figure. \ref{fig:RMSE_SNR} depicts the RMSE$(\mathbf{u})$ of the proposed hybrid TDOA/AOA-based algorithm alongside the square root of the CRLB as functions of $\rho$. The results confirm the effectiveness of the proposed estimator across various noise conditions.

Note that to the best of our knowledge, there is not any method to be able to locate a source in 3-D space using only one TDOA and one AOA measurement. Thus, we just compared performance of the proposed estimator with the CRLB, which sets a lower bound on variance of any unbiased estimator.

In the second scenario, we investigate the localization performance of a target traveling along the $x$-axis to assess the localization performance across various potential target locations within the surveillance area. The measurement noise parameters are set as $(\sigma_r,\sigma_\phi,\sigma_\theta)=(10\text{m}, 1^\circ, 1^\circ)$. The $y$ and $z$ coordinates of the target are kept fixed, as in the first scenario. The results, shown in Figure. \ref{fig:Target_sweep}, confirm that the proposed method consistently achieves performance that matches the CRLB for all target locations.

In the third scenario, the target's location is identical to that in the first scenario, while the measurement noise setup is configured as $(\sigma_r,\sigma_\phi,\sigma_\theta)=(10\text{m}, 1^\circ, 1^\circ)$.
In this scenario, we aim to evaluate the performance of the proposed method by comparing it with the Maximum Likelihood (ML) estimator which is implemented using the Gauss-Newton optimizer. While this method offers high accuracy, it is computationally intensive and requires initialization close to the true target position to ensure convergence. To provide a fair assessment of its accuracy, we initialized the ML estimator with the true source location. Figure. \ref{fig:CDF} illustrates the experimental cumulative distribution function (CDF) of positioning error for the proposed method and the ML estimator. As shown, the proposed algorithm achieves localization performance comparable to that of the ML estimator in this scenario.

\section{Conclusion} \label{sec:6}
This letter addresses the problem of single-source localization by utilizing TDOA and AOA measurements. A closed-form solution is derived by formulating pseudo-linear equations, with the target position subsequently estimated using a WLS approach. The efficacy of the proposed method is demonstrated through both analytical derivations and numerical simulations, showing that it achieves the CRLB under the assumption of small Gaussian measurement noise.

\bibliographystyle{IEEEtran}
\balance
\bibliography{refs}

\end{document}